\documentclass{article}

\usepackage{arxiv}

\usepackage[utf8]{inputenc} 
\usepackage[T1]{fontenc}    
\usepackage{hyperref}       
\usepackage{url}            
\usepackage{booktabs}       
\usepackage{amsfonts}       
\usepackage{nicefrac}       
\usepackage{microtype}      
\usepackage{lipsum}		
\usepackage{graphicx}
\usepackage{natbib}
\usepackage{doi}
\usepackage{verbatim}
\usepackage{subcaption}
\usepackage[misc]{ifsym}

\usepackage{algorithm2e}
\usepackage{amsmath}
\usepackage{array}
\usepackage{multirow}
\usepackage{subcaption}
\usepackage{listings}

\usepackage{pgf}
\usepackage{tikz}
\usetikzlibrary{arrows,automata}
\usetikzlibrary{positioning}

\tikzset{
    state/.style={
           rectangle,
           rounded corners,
           draw=black, very thick,
           minimum height=2em,
           minimum width=2em,
           inner sep=2pt,
           text centered,
           },
}
\usepackage{placeins}
\usepackage{amssymb}
\usepackage{enumitem}
\usepackage{pifont}

\title{Identifying Substitute and Complementary Products for Assortment Optimization with Cleora Embeddings}

\author{ 
    \href{https://orcid.org/
0000-0002-3434-6320}{\includegraphics[scale=0.06]{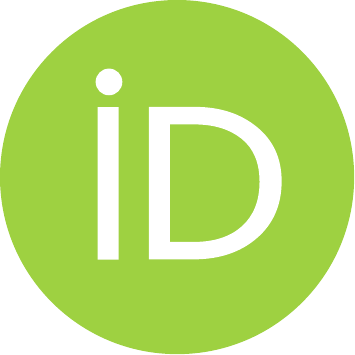}\hspace{1mm}Sergiy Tkachuk$^1$},
\href{https://orcid.org/0000-0002-3407-7570}{\includegraphics[scale=0.06]{orcid.pdf}\hspace{1mm}Anna Wr{\'o}blewska$^2$},
Jacek Dąbrowski$^{4}$,
\href{https://orcid.org/0000-0001-6716-610X}{\includegraphics[scale=0.06]{orcid.pdf}\hspace{1mm}Szymon {\L}ukasik$^{1,3}$}\\
 	$^1$Systems Research Institute,
		Polish Academy of Sciences\\
		ul.\ Newelska 6, 01-447 Warsaw, Poland\\
		Email: \texttt{\{stkachuk,slukasik\}@ibspan.waw.pl} \\
 	$^2$Faculty of Mathematics and Information Science, Warsaw University of Technology, Warsaw, Poland \\ Email: \texttt{\{anna.wroblewska1\}@pw.edu.pl} \\
	$^3$Faculty of Physics and Applied Computer Science,
		AGH University of Science and Technology\\
		al.\ Mickiewicza 30, 30-059 Krak\'{o}w, Poland\\
		Email: \texttt{slukasik@agh.edu.pl} \\
	$^4$Synerise,
		Warsaw, Poland\\
		Email: \texttt{jack.dabrowski@synerise.com} 
 }

\date{}

\renewcommand{\shorttitle}{Identifying Substitute and Complementary Products}

\hypersetup{
pdftitle={Identifying Substitute and Complementary Products for Assortment Optimization with Cleora Embeddings},
pdfsubject={cs.LG},
pdfauthor={Sergiy Tkachuk, Anna Wr{\'o}blewska, Jacek Dąbrowski, Szymon {\L}ukasik},
pdfkeywords={Substitutes, Complementary products, Recommendation systems, Assortment optimization Cleora, Graph embeddings},
}

\begin{document}
\maketitle

\begin{abstract}
Recent years brought an increasing interest in the application of machine learning algorithms in e-commerce, omnichannel marketing, and the sales industry. It is not only to the algorithmic advances but also to data availability, representing transactions, users, and background product information. Finding products related in different ways, i.e., substitutes and complements is essential for users' recommendations at the vendor's site and for the vendor - to perform efficient assortment optimization. 

The paper introduces a novel method for finding products' substitutes and complements based on the graph embedding Cleora algorithm. We also provide its experimental evaluation with regards to the state-of-the-art Shopper algorithm, studying the relevance of recommendations with surveys from industry experts. It is concluded that the new approach presented here offers suitable choices of recommended products, requiring a minimal amount of additional information. The algorithm can be used in various enterprises, effectively identifying substitute and complementary product options. 
\end{abstract}


\section{Introduction}

The recommendation systems' performance and impact demonstrate the importance of their implementation in the contemporary retail industry~\cite{hosanagar2014will}. Companies like Netflix \cite{bennett2007netflix}, Amazon \cite{smith2017two} or YouTube \cite{davidson2010youtube} utilize customers' data to influence their preference and choices using cognitive bias techniques like 'anchoring effect' \cite{adomavicius2013recommender}. As an integral part of the online retailing platforms, recommendation systems also impact the willingness-to-pay of users~\cite{adomavicius2012effects}.

The outstanding performance of recommendation engines does not stop researchers from developing it further. It is uncovered that modeling product relationships improve the accuracy of the recommendations~\cite{zhao2017improving}. Definition and modeling of substitutable and complementary links between products are of paramount importance for the next generation recommendation systems.

While recommendation systems have been experiencing superiority in the latest decade, the research on product relationships started even earlier \cite{mcalister1983identifying}. According to \cite{zhang2020} and \cite{mcauley2015inferring}, substitutable products are interchangeable and can be purchased instead of each other, e.g., one electronic component for another. Complementary products might be bought and used together; they experience joint demand, e.g., laptops and mice. In a typical recommendation system, both substitutes and complements may serve as good recommendation candidates. However, it depends on a purchase stage in which is a customer~\cite{zhang2020}. For example, when a user is evaluating a product (first stage of purchase), it is reasonable to recommend other substitutable products to better match their needs. However, it also is sensible to recommend complementary products (or even a bundle of products) but at further stages of the purchase process.

From the point of view of applying behavioral data, co-purchase and co-view relations can be used implicitly to represent complementarity and substitutionary recommendation strategies, respectively. It is, however, not always the case, and it is hard to use these implicit assumptions as a direct indication, especially for e-commerce vendors~\cite{zhang2016}. Therefore, it is crucial to building recommendation systems that can support decisions also from the retailer's perspective. 

This paper's primary goal is to introduce the new algorithm for identifying substitutes and complementary products using Cleora \cite{rychalska2021cleora} -- a general-purpose graph embedding algorithm. Besides describing the algorithm, the results of its expert-based evaluation and comparison with the selected state-of-the-art solution -- Shopper algorithm -- on real-world data are provided. We also identify the most challenging research gaps in gathering and defining substitutes and complementary products in massive omnichannel retail datasets. Our goal is to provide relevant recommendations which might support decisions such as assortment optimization, building marketing strategy, or simply finding missing product replacements for various industries. 

The paper is organized as follows. In the next Section, identifying substitutes and complementary products is reviewed from the perspective of existing techniques and available datasets -- which might be used for testing new algorithms. Section~\ref{sec:our-approach} describes the introduced algorithm. Next, in Section~\ref{sec:experimental-setup}, the experimental setup is covered in detail. Section~\ref{sec:results} overviews the experiments' results, while the last part of the paper (Section~\ref{sec:conclusions}) contains the final discussion and area of future research.

\section{Related Work} \label{sec:related-works}

Research on identifying product relationships in the recommendation systems space from the customer perspective is well-established. In this section, we briefly discuss the highlights of the literature review. Then we scheme the requirements for datasets to build and test recommendations for vendors reliably.

\subsection{Existing Algorithms}

It should be noted first that while the customer-facing recommendations are studied firmly, the research around recommendations for retail business performance is scarce. Consumers' preferences and subjective choices make it hard to assess what product is the suitable substitute or complement to a product of interest. Hence, it is challenging to decide on the relevant assortment or stocks at a given point of the sale, considering the constraints companies face (e.g., stocking costs or area of a store). We believe that customer-facing recommendation systems design and algorithms of identifying product links can be used in business support systems, such as autonomous assortment optimization.

With that respect, \cite{zheng2009substitutes} proposed tagging method employing a conventional collaborative filtering algorithm. The authors tagged category pairs manually and introduced a third 'Unsure' option for pairs with relationships that seemed uncertain. This additional dimension made the approach outstanding. However, the manual way of labeling is not sustainable. Given the dynamic nature of assortment planning and consumers' preferences, it requires significant effort to maintain and scale the solution. In the following study, \cite{mcauley2015inferring} introduced a different perspective to inferring substitutable and complementary products mining. The designed system called Sceptre can model and predict links between products from the text of their reviews and description. The data used for the Sceptre inference comes from Amazon and constitutes a large corpus of millions of products, including co-purchasing and browsing data. The work builds on top of the LDA \cite{blei2003latent} model family and aims to identify topics that explain document networks' links. Unlike the tagging method, the Sceptre assumes the substitutable and complementary products definition. Products that 'also viewed' are referred to as substitutes, whereas 'also bought' products are complementary. A similar notion of product links was also exploited by \cite{trofimov2018inferring}. Authors of \cite{wang2018path} proposed a path-constrained framework (PMSC) which is similar to, and arguably outperformed, the Sceptre~\cite{mcauley2015inferring}. The framework introduced a novel mapping function to project products embedding vectors into two separate spaces increasing the model's substitutable and complementary discriminative ability. Based on proprietary Walmart dataset \cite{xu2020product} introduced product knowledge graph (PKG) embedding approach for learning products relationships in e-commerce. Similarly, in \cite{mcauley2015inferring}, the authors took the same assumption concerning substitutes (co-view) and complementary products (co-buy) definitions. The study results demonstrate PKG out-performance over traditional knowledge graphs (KG). Another worth mentioning, the embedding-related approach is a semi-supervised Substitute Products Embedding Model (SPEM) \cite{zhang2019inferring}. The deep embedding method demonstrated superior results over a set of high-performing solutions, including Sceptre \cite{mcauley2015inferring}. The mentioned studies utilizes similar validation metrics -- Hits@k.

One of the most recent models called Shopper used as a reference point in our research was proposed by~\cite{ruiz2020shopper}. Shopper model can identify pairs of substitutes and complements -- where purchasing one item makes the second one less or more appealing, respectively. One of the model's key features is called "thinking ahead," where a customer considers the next choice in making the current choice. It makes the model more generally applicable for both online and offline shopping. Shopper uses the Bayesian model of market baskets in the methodological layer, estimating the items' latent features, seasonal effects, preferences of the customers, and price sensitivities to obtain substitutes and complements estimation. 

\subsection{Datasets}

As we mentioned above, the research around retail business performance recommendations is scarce, so do the adequate datasets. To have a possibility of sufficiently assessing the notion of substitution and complementing regarding e-commerce products, we need to have detailed information on products and transaction baskets available.

Requirements for the needed information are as follows:
\begin{enumerate}
\item \textbf{Transaction baskets.} Transactions have their identifiers. They should be gathered into baskets so that all the products purchased at one transaction should be identifiable. The quantity of each product in one transaction should also be given. 
\item \textbf{Precise product identification and differentiation.}  Exact information about individual products: EAN or other identifiers
\item \textbf{Exact product description.} Product name/title, precise information on the brand, attributes, e.g., capacity, size.
\item \textbf{Unified product attribute schema.} All input products should be prepared in the same way to avoid unnecessary work of re-description. 
\item \textbf{Seasonal information on products.} It can be derived from other data, e.g., transaction date. 
\item \textbf{Product price of purchase.} Price given with the purchase transaction (the prices can change along the time). 
\item \textbf{Purchases in shops should be distinguishable.}  Each shop should have an identifier, the transaction should have a shop (when purchased) identification. 
\item \textbf{Clients should have their unique identifiers.} They should be distinguished from others.  Usually, algorithms need to gather purchases of each client.  
\end{enumerate}
The last two requirements are not essential for algorithms to produce their output results but can be very useful, depending on an algorithm's options. 

Many e-commerce datasets are publicly available on the Kaggle platform and the Web (see~\cite{lionbridge_datasets}). However, very few contain all the required precise information.
One of the massive e-commerce product datasets is the Amazon reviews dataset~\cite{amazon_dataset} that contains product details and the relations between products bought together. However, there is no information about which products were bought in one basket (in one purchase transaction), required in Shopper and our Cleora settings. The latest extensive dataset with precise information on products and their attributes is given in~\cite{wdc2020}. However, there are also no required product relation links or even behavioral customer patterns to infer the links. On the other hand, datasets containing more precise information about customer behaviors and transaction baskets have tiny details on products themselves, e.g., the Dunnhumby's Complete Journey dataset~\cite{dunnhumbys}, and others~\cite{brazilian_dataset,rocket_dataset}. Without product names and their precise attributes, it seems impossible to assess if given substitutions or complementations are suitable.

\section{Proposed Approach}
\label{sec:our-approach}

The algorithm introduced here uses the embedding of graphs representing product relationships. By embedding products in the latent space, it is possible to present proximities between products. Thus, products situated close to each other are identified as related.

A variety of embedding methods allow the transformation of different modalities of input data into a vector space. Depending on modality characteristics, a wide range of transformation procedures can be employed, yielding sparse representations (e.g. one-hot encoding of text~\cite{Cerda_jml_2018}) or dense vector representations, e.g. language modelling (e.g. word2vec~\cite{mikolov2013efficient}, gloVe~\cite{Pennington_emnlp_2014}, BERT~\cite{Devlin_naacl_2019}), node2vec results on graph representations~\cite{Grover_kdd_2019}, image deep neural networks results (e.g. VGG, ResNet embeddings)~\cite{Kornblith_cvpr_2019}. Here a new hypergraph embedding approach -- Cleora~\cite{rychalska2021cleora} is being used. In the experiments, it was proved to be not only accurate but also highly efficient for large datasets. Our approach will be covered in detail in the following part of this section.

The graph format is extremely limited, allowing only for links between two entities. On the other hand, hypergraphs allow representing more extensive relations. For instance, datasets of shopping baskets or user's online sessions form natural hypergraphs. All the products in a shopping basket are together bound by the relation of belonging to a single basket. All the items viewed during an online session are strongly coupled as well. Furthermore, the number of items in both can be significantly larger than two. Hence, a single basket is a hyperedge.

It has been empirically confirmed~\cite{rychalska2021cleora} that the hypergraph approach works much better than graphs for product embeddings, website embeddings, and similar data. It is more expressive and captures the underlying data generating process better.

In general a graph $G$ constitutes a pair $(V, E)$ where $V$ denotes a set of vertices and $E \subseteq (V \times V)$ -- edges connecting them. Product relationships will be represented by undirected graphs, where an edge is an unordered pair of nodes. 
An embedding of a graph $G=(V,E)$ is denoted as  matrix $T$, with size $|V| \times d$,  where $d$ is the dimension of the embedding. The i-th row of matrix $T$ (denoted as $T_{i,*}$) corresponds to a node $i \in V$ while $T_{i,j}$, $j \in \{1, ..., d\}$, captures a different feature of node $i$.

The input transactional data were grouped by order or invoice concatenating product codes (counterpart of EAN number). The data is not sequential, i.e. order of products on the transaction list is not meaningful. An example of the input data shape for sample products p1-p6 is illustrated in Table~\ref{tab:cleora-input-data}.
\begin{table}[!htb]
\centering
\begin{tabular}{l|lll}
            & Product Basket \\
        \hline
Transaction 1  & p1    p3    p4   \\
Transaction 2  & p2    p4   \\
Transaction 3  & p5    p6    p3   \\          
\end{tabular}
\caption{Example of the input data shape.}
\label{tab:cleora-input-data}
\end{table}

The algorithm used for creating product embeddings is enclosed as Algorithm \ref{algo-cleora}. Its implementation created in Rust, with Python bindings and compiled binaries is available as open-source \cite{cleoracode}.

\begin{algorithm}[!htb]
 \KwData{Graph $G = (V,E)$ with set of nodes $V$ and set of edges $E$, iteration number $I$, chunk number $Q$, embedding dimensionality $d$} 
 \KwResult{Embedding matrix  $\mathbf{T} \in \mathbb{R}^{|V| \times d}$ }
Divide graph $G$ into $Q$ chunks. Let $G_{q}$ be the $q$-th chunk with edge set $E_q$\;

\For{q from 1 to Q} {
 For graph chunk $G_{q}$ compute random walk transition matrix $\mathbf{M}_{q}=\frac{e_{a b}}{deg(v_{a})}$ for $a b \in E_q$, where $e_{ab}$ is the number of edges running from node $a$ to $b$  \;
 Initialize chunk embedding matrix $\mathbf{T}^{q}_{0} \sim U\left(-1, 1\right)$ \;
 \For{i from 1 to I}{
    $\mathbf{T}^{q}_{i}=\mathbf{M}_{q} \cdot \mathbf{T}^{q}_{i-1}$\;
    $L_2$ normalize rows of $\mathbf{T}^{q}_{i}$\;
 }
 $\mathbf{T}^{q}$ = $\mathbf{T}^{q}_{I}$ \;
 }
 
 \For{v from 1 to $|V|$} {
    $\mathbf{T_{v,*}}$ = $\sum_{q=1}^{Q}{w_{q,v} \times \mathbf{T^{q}_{v,*}}}$ where $w_{q,v}$ is the node weight $w_{q,v} = \frac{|n \in V_{q}: n = n_v |}{\sum_{k=1}^{Q}{|n \in V_{k}: n = n_v |}}$
 }
 \caption{Cleora algorithm as introduced in~\cite{rychalska2021cleora}}
 \label{algo-cleora}
\end{algorithm}

All the embeddings were trained with one and six iterations to reflect complementary and substitutable product relationships, respectively, with the same dimensionality (1024). The closest neighbor for embeddings obtained with one Cleora iteration is potentially the best complementary product. Whereas the one for embeddings obtained with six Cleora iteration -- when the algorithm is known to converge \cite{rychalska2021cleora} -- is most likely to be a substitute item. The nearest neighbors for each product were calculated using cosine similarity, which is commonly used to compare other kinds of embeddings, e.g. text-based embeddings.

The algorithm is executed in a feasible time, even for large datasets. This efficiency is achieved by combining two major kinds of vertex embedding methods.
\begin{enumerate}
    \item \textbf{Random walks.} These approaches take the input graph and stochastically generate "paths" (sequences of vertices) from a random walk on the graph. As if a bug was hopping from one vertex to another, along the edges. They are trained with stochastic gradient descent applied to various objectives, most of them similar to skip-gram or CBOW known from word2vec~\cite{mikolov2013efficient} and can differ slightly in how the random walks are defined.
    \item \textbf{Graph Convolutional Networks.} These models treat the graph as an adjacency matrix, define a "neighborhood mixing" operator called a graph convolution, and some non-linearities in between. They can be trained with SGD applied to various objectives as well.
\end{enumerate}

Finally, it is essential to highlight once again that Cleora code is open-source which allows reproducing our experiments and modify them relatively quickly and easily.

\section{Experimental Setup}
\label{sec:experimental-setup}

In our tests, the Cleora algorithm was benchmarked against Shopper~\cite{ruiz2020shopper}.  The Shopper was selected amongst other algorithms as its "thinking ahead" property suits the best offline shopping behavior in our research elaboration frame. Similarly to Cleora~\cite{rychalska2021cleora}, it generates product vectors reflecting assortment categories relationships. Moreover, Shopper is similar to contemporary recommendation systems that utilize matrix factorization techniques, making it appealing to explore from the business applicability and relevance standpoint. Given the research's collaborative academic and commercial nature, the empirical study described in the Shopper paper coincided with a progressing project for one of the largest European retail chains performed by Synerise, which impacted the benchmark model's choice.

As mentioned in the Related Datasets Section~\ref{sec:related-works}, the required data to reliably assess the interested product relations, i.e., supplementary and complementary products, are not publicly available. In Table~\ref{table:dataset-requirements}, we gathered the required dataset features and the particular needs for the chosen algorithms (Cleora and Shopper). Generally, we need data that are commonly included in purchase receipts. Additionally, we can use data that allow collecting all client purchases for more extended variants of the algorithms. It is vital not to use sequential transaction data because they are biased by online recommender system suggestions and other online and offline marketing strategies.

\begin{table*}[!htb]
\centering
\caption{Requirements for the dataset to assess product similarity or relatedness (supplementary, complementary products). Note: \checkmark -- requirement should be fulfilled, "--" -- requirement may not be fulfilled, (\checkmark) -- requirement may be fulfilled for the extended variant of the method. }
\label{table:dataset-requirements}
\begin{tabular}{p{0.3cm}p{8.4cm}p{1cm}p{1cm}}
\hline
No & Requirement name   & Cleora & Shopper \\ \hline
1 & \textbf{Transaction baskets} & \checkmark & \checkmark \\
2 & \textbf{Precise product identification and differentiation}  & \checkmark & \checkmark \\ 
3 & \textbf{Exact product description} & -- & -- \\ 
4 & \textbf{Unified product attribute schema} & (\checkmark) & (\checkmark) \\ 
5 & \textbf{Seasonal information on products} & -- & \checkmark \\
6 & \textbf{Product price of purchase} & -- & \checkmark \\ 
7 & \textbf{Purchases in shops should be distinguishable} & (\checkmark) & (\checkmark) \\ 
8 & \textbf{Client should have their unique identifiers} &  (\checkmark) & (\checkmark) \\ 
\hline
\end{tabular}
\end{table*}

We gathered our input dataset from the mentioned big retail chain in Europe. The dataset comprises about 561,135 products. As the input for substitution/complementary products identification we have selected products from three food categories, i.e., beers, spices, and yoghurts, covering 5,906 products. The dataset contains 351,422 transactions collected within 578 days for 3,634 customers. This data, grouped into baskets, was utilized as inputs to selection algorithms of substitutes and complements. Along with Cleora and Shopper, we have also used randomized generation of related products -- to serve as a baseline. The algorithms were used to generate two closest substitutes and two closest complementary products. 

Having these output products, we organized questionnaires for marketing industry specialists to assess the notion of supplementary and complementary links between one given focal product and their related substitutes and complementaries generated with different algorithms. It is worth mentioning that in contrast to \cite{mcauley2015inferring} we used experts judgment to validate models performance instead of Mechnical Turk -- to obtain reliable evaluation of the proposed technique. 

For the Cleora approach, we have not made any changes to the procedure's default scheme described in Section~\ref{sec:our-approach} using 1024 as the dimensionality of the embedding and one iteration for complementary products generation, and six -- for substitutes. Shopper algorithm was used with 
$K=100$, $lookahead=0$ and $max-iterations=1200$. The full set of parameters used for this algorithm was enclosed in the Appendix. The resulting items (substitutes and complements) for the Random algorithm were randomly selected from the entire retail chain assortment.

\section{Results}
\label{sec:results}

After preparing our suggestions for supplementary and complementary products using Cleora, Shopper, and random algorithms, we gathered human assessments. We prepared questionnaires for retail experts. The survey participants were asked to point to (1) two substitutes out of six proposed (two generated by each of the tested algorithms) and (2) two complementary products out of six offered (again, generated with the three tested algorithms). The survey covered 20 randomly selected products from each product class: beers, yoghurts, and spices. The answers were ordered randomly for each participant, and they did not know the source algorithm (that one generated the product suggestions). The respondents were also allowed to point to only one substitute/complementary product in each question. 

Altogether, 5,520 answers regarding substitutes and complementary products were recorded during the survey, with 138 questionnaires being completed (each having 20 questions). This substantial experimental data will be discussed in the following paragraphs of this Section. 

First, let us compare how both of the substitutes/complementary products were evaluated by the retail experts. Table \ref{tab:recommendation-substitutes-accuracies} provides ratios of products for which tested algorithms provided two substitutes indicated by experts regardless of their order (correct or reversed ones). 

\begin{table}[!htb]
\centering
\begin{tabular}{l|lll}
          & Beers & Yoghurts & Spices \\ \hline
Cleora    & 0.60      & 0.76       & 0.77    \\
Shopper   & 0.05      & 0.00       & 0.01    \\
Random    & 0.00      & 0.00       & 0.00    \\          
\end{tabular}
\caption{Accuracy of the recommendation algorithms for substitutes in three food product categories.}
\label{tab:recommendation-substitutes-accuracies}
\end{table}

In all cases, the Cleora algorithm significantly outperforms both the Shopper algorithm and random selection of substitutes. The accuracy of our proposed Cleora-based techniques is at least ten times higher than the benchmarked ones. 

Similar comparison for complementary products was enclosed in Table~\ref{tab:recommendation-complements-accuracies}. Again our approach proves to be the most competitive, with the Shopper algorithm being slightly more accurate only for the "yoghurt" product category.  It also observed that finding complementary products is more difficult and subjective, with the highest accuracy just over 50\% for Cleora and the "spices" category. 

\begin{table}[!htb]
\centering
\begin{tabular}{l|lll}
        & Beers & Yoghurts & Spices  \\
        \hline
Cleora    & 0.15 & 0.19    & 0.55              \\
Shopper   & 0.09 & 0.21    & 0.05     \\
Random    & 0.00    & 0.00       & 0.00      
\end{tabular}
\caption{Accuracy of the recommendation algorithms for complements in three food product categories.}
\label{tab:recommendation-complements-accuracies}
\end{table}

\begin{figure}
\centering
\includegraphics[width=0.45\textwidth]{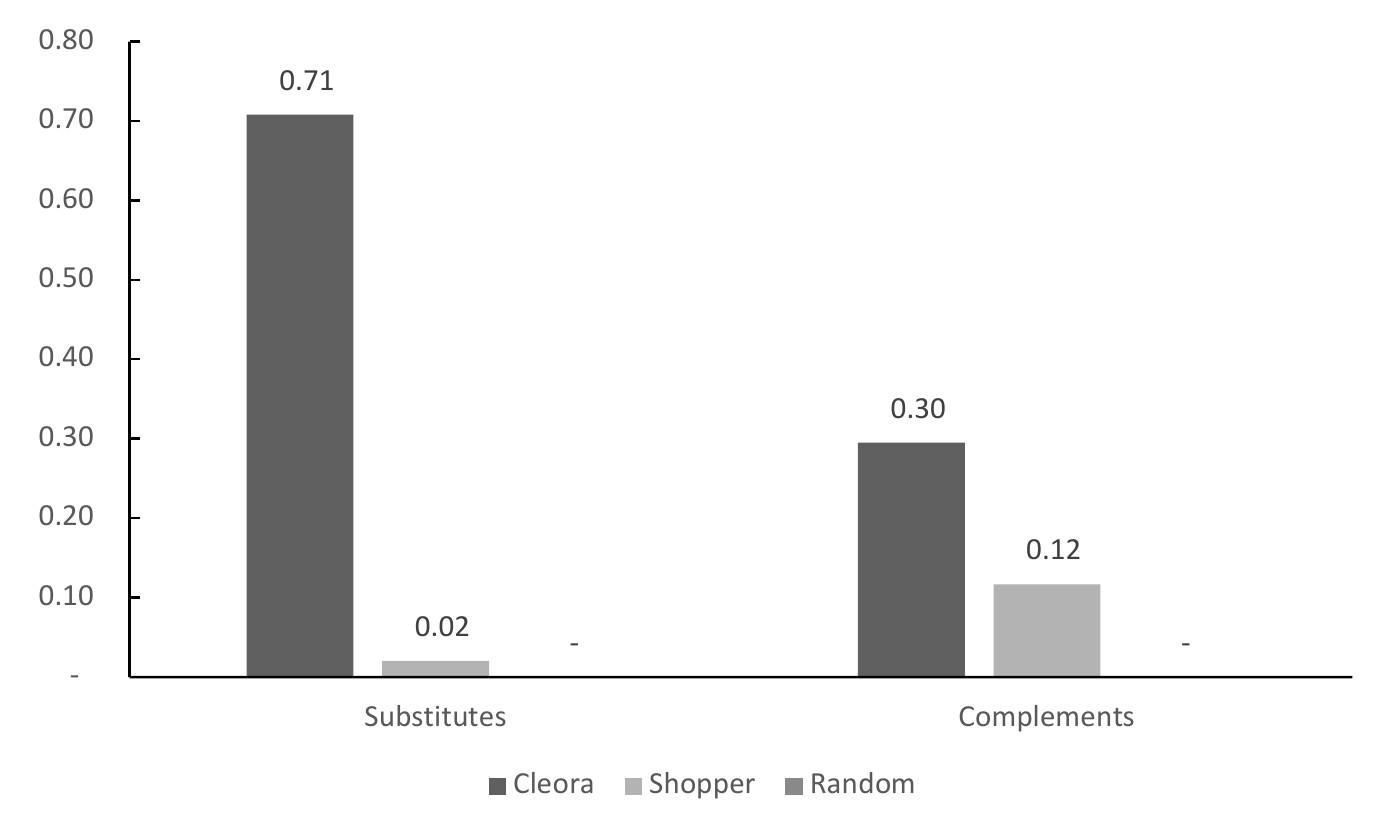}
\caption{Weighted accuracy for evaluated algorithms taking into account product class representation in survey data. Note: Random results are close to zero.} \label{fig:accuracy}
\end{figure}

Figure~\ref{fig:accuracy} summarizes this part of the study's results. It demonstrates weighted mean values, with weights attributed to each of the product classes corresponding to each class's number of surveys, i.e.:
\begin{equation}
acc=\frac{\sum acc_i * m_i}{m}    
\end{equation}
with $acc_i$ representing accuracy for product category $i$, $m_i$ indicating the number of answers recorded for that category, and $m$ -- the overall number of answers. It can be again seen that the proposed approach based on Cleora is superior to the state-of-art Shopper algorithm. 

Tables \ref{tab:substorder} and \ref{tab:complorder} contain the number of substitutes and complementary product pairs, which were pointed out by each algorithm. This time we study the correctness of the order of this indication with regards to the order indicated by experts. It can be seen that Cleora-based generation of related products is much more predictable in terms of the proper order of substitutes/complements. 

\begin{table*}[!htb]
\resizebox{\textwidth}{!}{
\begin{tabular}{ll|lll|lll|lll|}
\multicolumn{2}{l|}{Number   of substitutes} & \multicolumn{3}{c|}{500}              & \multicolumn{3}{c|}{400}              & \multicolumn{3}{c|}{480}              \\
\hline
  &           & \multicolumn{3}{c|}{Beers}            & \multicolumn{3}{c|}{Yoghurts}         & \multicolumn{3}{c|}{Spices}           \\ \hline
  &           & Correct order & Reversed order & Sum & Correct order & Reversed order & Sum & Correct order & Reversed order & Sum \\
\multirow{3}{*}{}    & Cleora    & 164           & 134            & 298 & 160           & 143            & 303            & 208           & 162            & 370 \\
 & Shopper   & 4             & 19             & 23  & 1             & -              & 1   & 1             & 5              & 6   \\
 & Random    & -             & -              & -   & -             & -              & -   & -             & -              & -  \\ \hline
\end{tabular}
}
\caption{Pairs of substitutes correctly identified in exact order (as indicated by the algorithms) or reversed order.}
\label{tab:substorder}
\end{table*}

\begin{table*}[!htb]
\resizebox{\textwidth}{!}{
\begin{tabular}{ll|lll|lll|lll|}
\multicolumn{2}{l|}{Number   of complements} & \multicolumn{3}{c|}{500}              & \multicolumn{3}{c|}{400}              & \multicolumn{3}{c|}{480}              \\ \hline
 &           & \multicolumn{3}{c|}{Beers}            & \multicolumn{3}{c|}{Yoghurts}         & \multicolumn{3}{c|}{Spices}           \\ \hline
 &           & Correct order & Reversed order & Sum & Correct order & Reversed order & Sum & Correct order & Reversed order & Sum \\
\multirow{3}{*}{Recommender}    & Cleora    & 28            & 46             & 74  & 45            & 29             & 74  & 170           & 95             & 265 \\
& Shopper   & 6             & 39             & 45  & 60            & 24             & 84  & 15            & 9              & 24  \\
 & Random    & 1             & 4              & 5   & 1             & 1              & 2   & -             & 1              & 1 \\ \hline
\end{tabular}
}
\caption{Pairs of complementary products correctly identified in exact order (as indicated by the algorithms) or reversed order.}
\label{tab:complorder}
\end{table*}

To investigate the algorithm performance, it is also important to evaluate if the algorithm successfully provides an accurate first substitute/complementary product. The comparison for this aspect of the considered problem was provided in Tables \ref{tab:subst_first} and \ref{tab:compl_first}. 

\begin{table}[!htb]
\centering
\begin{tabular}{l|lll}
        & Beers & Yoghurts & Spices  \\
        \hline
Cleora  & 38.9\% & 45.1\%    & 43.2\%  \\
Shopper & 6.4\%  & 1.3\%     & 3.6\%   \\
Random  & 4.3\%  & 1.3\%     & 0.4\%   \\ 
\end{tabular}
\caption{Percentage of the substitute products given by experts which include the first substitute generated by each algorithm.}
\label{tab:subst_first}
\end{table}

\begin{table}[!htb]
\centering
\begin{tabular}{l|lll}
        & Beers & Yoghurts & Spices  \\
        \hline
Cleora  & 17.1\% & 22.4\% & 40.4\% \\
Shopper & 8.2\%  & 26.8\% & 8.6\%  \\
Random  & 7.4\%  & 5.3\%  & 3.9\%
\end{tabular}
\caption{Percentage of the complementary products given by experts which include the first complement generated by each algorithm.}
\label{tab:compl_first}
\end{table}

It was established that the first substitute generated with Cleora was an excellent choice in the experts' opinion. In the case of complementary products, similar findings were derived. Only in the "yoghurt" product class, the result obtained with Shopper was slightly more competitive. 

\section{Conclusions}
\label{sec:conclusions}

The paper introduced a new method of generating substitutes and complementary products for a given item, using unordered transactional information. Thanks to its minimal data requirements, it can be used for offline and online assortment recommendations. The algorithm was demonstrated here to be very accurate.  It has outperformed the state-of-art Shopper algorithm for the same testing data, which was confirmed by a significant sample of experts' decisions used as a point of reference (gold standard answers).

It is essential to stress that the approach proposed here is highly efficient computationally and involves solely CPU-based methods. Cleora embeddings for both substitutes and complementary products were trained in less than thirty minutes each, while latent vectors for the benchmark algorithm -- Shopper -- required more than 24 hours of GPU-based and CPU hyper-threading computations. Thus, the presented algorithm allows to quickly and efficiently identify substitution and complementary products, which can be very helpful in vendors' settings having to move or change their stock products -- for their assortment optimization.

Additionally, the obtained representations of latent product features (from Cleora) can be utilized to perform behavioral segmentation of products. It allows recreating links between individual items and clusters of products being purchased in a similar context. Moreover, in similar settings, Cleora can be exploited to segment customers based on their behavioral patterns. The tool is independent of the industry for which the substitute/complementary products problem is being studied, and it uses only the context of product purchase/order, thus it can be used also for related tasks such as suggestion of parts to order etc.

\section*{Acknowledgements}
This work was co-financed from the project "Development of inference technology and prediction of consumer behavior for the automation of marketing and sales processes" co-financed by the European Regional Development Fund under the Regional Operational Program of the Małopolska Region for 2014-2020. It was also partially financed (supported) by the Faculty of Physics and Applied Computer Science AGH UST statutory tasks within subsidy of Ministry of Science and Higher Education.

Research was also partially funded by the Centre for Priority Research Area Artificial Intelligence and Robotics of Warsaw University of Technology within the Excellence Initiative: Research University (IDUB) programme (grant no 1820/27/Z01/POB2/2021).

\appendix
\section{Shopper algorithms parameters set}
For Shopper algorithm the following input parameters -- derived from the existing studies -- were used:

\noindent\texttt{+K=100 +Kgroup=0 +fixKgroup=0 +seed=0 +rfreq=200000 +saveCycle=5000 +max-iterations=1200 +negsamples=50 +nsFreq=-1 +likelihood=1 +lookahead=0 +avgContext=1 +symmetricRho=0 +checkout=1 +shuffle=1 +zeroFactor=0.100000 +batchsize=5000 +userVec=3 +itemIntercept=1 +price=10 +day=10 +normPrice=0 +normPriceMin=0 +step\_schedule=2 +eta=0.010000 +gamma=0.900000 +valTolerance=0.000001 +valConsecutive=5 +keepOnly=-1 +keepAbove=-1 +thr\_llh=-100000.000000 +threads=16}

\noindent For initialization:

\noindent\texttt{+stdIni=0.100000 +iniPath= +iniFromGroup=}

\noindent Hyperparameters:

\noindent\texttt{+s2rho=1.000000 +s2alpha=1.000000 +s2theta=1.000000 +s2lambda=1.000000 +rtegamma=1000.000000 +shpgamma=100.000000 +rtebeta=1000.000000 +shpbeta=100.000000 +s2delta=0.010000 +s2mu=0.010000}

\FloatBarrier

\bibliographystyle{unsrtnat}
\bibliography{references}  

\begin{thebibliography}{32}
\providecommand{\natexlab}[1]{#1}
\providecommand{\url}[1]{\texttt{#1}}
\expandafter\ifx\csname urlstyle\endcsname\relax
  \providecommand{\doi}[1]{doi: #1}\else
  \providecommand{\doi}{doi: \begingroup \urlstyle{rm}\Url}\fi

\bibitem[Hosanagar et~al.(2014)Hosanagar, Fleder, Lee, and
  Buja]{hosanagar2014will}
Kartik Hosanagar, Daniel Fleder, Dokyun Lee, and Andreas Buja.
\newblock Will the global village fracture into tribes? recommender systems and
  their effects on consumer fragmentation.
\newblock \emph{Management Science}, 60\penalty0 (4):\penalty0 805--823, 2014.

\bibitem[Bennett et~al.(2007)Bennett, Lanning, et~al.]{bennett2007netflix}
James Bennett, Stan Lanning, et~al.
\newblock The netflix prize.
\newblock In \emph{Proceedings of KDD cup and workshop}, volume 2007, page~35.
  Citeseer, 2007.

\bibitem[Smith and Linden(2017)]{smith2017two}
Brent Smith and Greg Linden.
\newblock Two decades of recommender systems at amazon. com.
\newblock \emph{Ieee internet computing}, 21\penalty0 (3):\penalty0 12--18,
  2017.

\bibitem[Davidson et~al.(2010)Davidson, Liebald, Liu, Nandy, Van~Vleet, Gargi,
  Gupta, He, Lambert, Livingston, et~al.]{davidson2010youtube}
James Davidson, Benjamin Liebald, Junning Liu, Palash Nandy, Taylor Van~Vleet,
  Ullas Gargi, Sujoy Gupta, Yu~He, Mike Lambert, Blake Livingston, et~al.
\newblock The youtube video recommendation system.
\newblock In \emph{Proceedings of the fourth ACM conference on Recommender
  systems}, pages 293--296, 2010.

\bibitem[Adomavicius et~al.(2013)Adomavicius, Bockstedt, Curley, and
  Zhang]{adomavicius2013recommender}
Gediminas Adomavicius, Jesse~C Bockstedt, Shawn~P Curley, and Jingjing Zhang.
\newblock Do recommender systems manipulate consumer preferences? a study of
  anchoring effects.
\newblock \emph{Information Systems Research}, 24\penalty0 (4):\penalty0
  956--975, 2013.

\bibitem[Adomavicius et~al.(2012)Adomavicius, Bockstedt, Curley, and
  Zhang]{adomavicius2012effects}
Gediminas Adomavicius, Jesse~C Bockstedt, Shawn~P Curley, and Jingjing Zhang.
\newblock Effects of online recommendations on consumers' willingness to pay.
\newblock In \emph{Decisions@ RecSys}, pages 40--45, 2012.

\bibitem[Zhao et~al.(2017)Zhao, McAuley, Li, and King]{zhao2017improving}
Tong Zhao, Julian McAuley, Mengya Li, and Irwin King.
\newblock Improving recommendation accuracy using networks of substitutable and
  complementary products.
\newblock In \emph{2017 International Joint Conference on Neural Networks
  (IJCNN)}, pages 3649--3655. IEEE, 2017.

\bibitem[McAlister and Lattin(1983)]{mcalister1983identifying}
Leigh McAlister and James~M Lattin.
\newblock \emph{Identifying substitute and complementary relationships revealed
  by consumer variety seeking behavior}.
\newblock Cambridge, Mass.: The Marketing Center, Massachusetts Institute
  of~Technology, 1983.

\bibitem[Zhang and Bockstedt(2020)]{zhang2020}
Mingyue Zhang and Jesse Bockstedt.
\newblock Complements and substitutes in online product recommendations: The
  differential effects on consumers’ willingness to pay.
\newblock \emph{Information \& Management}, 57\penalty0 (6):\penalty0 103341,
  2020.
\newblock ISSN 0378-7206.
\newblock \doi{https://doi.org/10.1016/j.im.2020.103341}.
\newblock URL
  \url{https://www.sciencedirect.com/science/article/pii/S0378720620302792}.

\bibitem[McAuley et~al.(2015)McAuley, Pandey, and
  Leskovec]{mcauley2015inferring}
Julian McAuley, Rahul Pandey, and Jure Leskovec.
\newblock Inferring networks of substitutable and complementary products.
\newblock In \emph{Proceedings of the 21th ACM SIGKDD international conference
  on knowledge discovery and data mining}, pages 785--794, 2015.

\bibitem[Zhang and Bockstedt(2016)]{zhang2016}
Mingyue Zhang and {Jesse C} Bockstedt.
\newblock Complements and substitutes in product recommendations: The
  differential effects on consumers' willingness-to-pay.
\newblock \emph{CEUR Workshop Proceedings}, 1679:\penalty0 36--43, 2016.
\newblock ISSN 1613-0073.

\bibitem[Rychalska et~al.(2021)Rychalska, Babel, Goluchowski, Michalowski, and
  Dabrowski]{rychalska2021cleora}
Barbara Rychalska, Piotr Babel, Konrad Goluchowski, Andrzej Michalowski, and
  Jacek Dabrowski.
\newblock Cleora: A simple, strong and scalable graph embedding scheme, 2021.

\bibitem[Zheng et~al.(2009)Zheng, Wu, Niu, and Bolivar]{zheng2009substitutes}
Jiaqian Zheng, Xiaoyuan Wu, Junyu Niu, and Alvaro Bolivar.
\newblock Substitutes or complements: another step forward in recommendations.
\newblock In \emph{Proceedings of the 10th ACM conference on Electronic
  commerce}, pages 139--146, 2009.

\bibitem[Blei et~al.(2003)Blei, Ng, and Jordan]{blei2003latent}
David~M Blei, Andrew~Y Ng, and Michael~I Jordan.
\newblock Latent dirichlet allocation.
\newblock \emph{the Journal of machine Learning research}, 3:\penalty0
  993--1022, 2003.

\bibitem[Trofimov(2018)]{trofimov2018inferring}
Ilya Trofimov.
\newblock Inferring complementary products from baskets and browsing sessions.
\newblock \emph{arXiv preprint arXiv:1809.09621}, 2018.

\bibitem[Wang et~al.(2018)Wang, Jiang, Ren, Tang, and Yin]{wang2018path}
Zihan Wang, Ziheng Jiang, Zhaochun Ren, Jiliang Tang, and Dawei Yin.
\newblock A path-constrained framework for discriminating substitutable and
  complementary products in e-commerce.
\newblock In \emph{Proceedings of the Eleventh ACM International Conference on
  Web Search and Data Mining}, pages 619--627, 2018.

\bibitem[Xu et~al.(2020)Xu, Ruan, Korpeoglu, Kumar, and Achan]{xu2020product}
Da~Xu, Chuanwei Ruan, Evren Korpeoglu, Sushant Kumar, and Kannan Achan.
\newblock Product knowledge graph embedding for e-commerce.
\newblock In \emph{Proceedings of the 13th international conference on web
  search and data mining}, pages 672--680, 2020.

\bibitem[Zhang et~al.(2019)Zhang, Yin, Wang, Chen, Chen, and
  Nguyen]{zhang2019inferring}
Shijie Zhang, Hongzhi Yin, Qinyong Wang, Tong Chen, Hongxu Chen, and Quoc
  Viet~Hung Nguyen.
\newblock Inferring substitutable products with deep network embedding.
\newblock In \emph{IJCAI}, pages 4306--4312, 2019.

\bibitem[Ruiz et~al.(2020)Ruiz, Athey, Blei, et~al.]{ruiz2020shopper}
Francisco~JR Ruiz, Susan Athey, David~M Blei, et~al.
\newblock Shopper: A probabilistic model of consumer choice with substitutes
  and complements.
\newblock \emph{Annals of Applied Statistics}, 14\penalty0 (1):\penalty0 1--27,
  2020.

\bibitem[Alex~Nguyen(2020)]{lionbridge_datasets}
LionBridge~AI Alex~Nguyen.
\newblock 24 best retail, sales, and ecommerce datasets for machine learning,
  2020.
\newblock URL
  \url{https://lionbridge.ai/datasets/24-best-ecommerce-retail-datasets-for-machine-learning/}.

\bibitem[He and McAuley(2016)]{amazon_dataset}
Ruining He and Julian McAuley.
\newblock Ups and downs: Modeling the visual evolution of fashion trends with
  one-class collaborative filtering.
\newblock In \emph{Proceedings of the 25th International Conference on World
  Wide Web}, WWW '16, page 507–517, Republic and Canton of Geneva, CHE, 2016.
  International World Wide Web Conferences Steering Committee.
\newblock ISBN 9781450341431.
\newblock \doi{10.1145/2872427.2883037}.
\newblock URL \url{https://doi.org/10.1145/2872427.2883037}.

\bibitem[Zhang et~al.(2020)Zhang, Bizer, Peeters, and Primpeli]{wdc2020}
Ziqi Zhang, Christian Bizer, Ralph Peeters, and Anna Primpeli.
\newblock Mwpd2020: Semantic web challenge on mining the web of html-embedded
  product data.
\newblock In Ziqi Zhang, editor, \emph{MWPD 2020 : Proceedings of the Semantic
  Web Challenge on Mining the Web of HTML-embedded Product Data co-located with
  the 19th International Semantic Web Conference (ISWC 2020) Athens, Greece,
  November 5, 2020}, volume 2720, pages 2--18, Aachen, November 2020. RWTH.
\newblock URL
  \url{https://madoc.bib.uni-mannheim.de/57428/,http://webdatacommons.org/largescaleproductcorpus/v2/index.html}.

\bibitem[Dunnhumby(2020)]{dunnhumbys}
Dunnhumby.
\newblock The complete journey dataset, 2020.
\newblock URL \url{https://www.dunnhumby.com/source-files/,
  https://www.kaggle.com/frtgnn/dunnhumby-the-complete-journey?select=product.csv}.

\bibitem[Olist(2019)]{brazilian_dataset}
Olist.
\newblock Brazilian e-commerce public dataset, 2019.
\newblock URL \url{https://www.kaggle.com/olistbr/brazilian-ecommerce}.

\bibitem[Retailrocket(2017)]{rocket_dataset}
Retailrocket.
\newblock Retailrocket recommender system dataset, 2017.
\newblock URL \url{https://www.kaggle.com/retailrocket/ecommerce-dataset}.

\bibitem[Cerda et~al.(2018)Cerda, Varoquaux, and Kégl]{Cerda_jml_2018}
Patricio Cerda, Gaël Varoquaux, and Balázs Kégl.
\newblock Similarity encoding for learning with dirty categorical variables.
\newblock \emph{Machine Learning}, 107\penalty0 (8):\penalty0 1477--1494,
  September 2018.
\newblock ISSN 1573-0565.
\newblock \doi{10.1007/s10994-018-5724-2}.
\newblock URL \url{https://doi.org/10.1007/s10994-018-5724-2}.

\bibitem[Mikolov et~al.(2013)Mikolov, Chen, Corrado, and
  Dean]{mikolov2013efficient}
Tomas Mikolov, Kai Chen, Greg Corrado, and Jeffrey Dean.
\newblock Efficient estimation of word representations in vector space.
\newblock \emph{arXiv preprint arXiv:1301.3781}, 2013.

\bibitem[Pennington et~al.(2014)Pennington, Socher, and
  Manning]{Pennington_emnlp_2014}
Jeffrey Pennington, Richard Socher, and Christopher~D Manning.
\newblock Glove: Global vectors for word representation.
\newblock In \emph{Proceedings of the 2014 conference on empirical methods in
  natural language processing (EMNLP)}, pages 1532--1543, 2014.

\bibitem[Devlin et~al.(2019)Devlin, Chang, Lee, and
  Toutanova]{Devlin_naacl_2019}
Jacob Devlin, Ming-Wei Chang, Kenton Lee, and Kristina Toutanova.
\newblock {BERT}: Pre-training of deep bidirectional transformers for language
  understanding.
\newblock In \emph{Proceedings of the 2019 Conference of the North {A}merican
  Chapter of the Association for Computational Linguistics: Human Language
  Technologies, Volume 1 (Long and Short Papers)}, pages 4171--4186,
  Minneapolis, Minnesota, June 2019. Association for Computational Linguistics.
\newblock \doi{10.18653/v1/N19-1423}.
\newblock URL \url{https://www.aclweb.org/anthology/N19-1423}.

\bibitem[Grover and Leskovec(2016)]{Grover_kdd_2019}
Aditya Grover and Jure Leskovec.
\newblock Node2vec: Scalable feature learning for networks.
\newblock In \emph{Proceedings of the 22nd ACM SIGKDD International Conference
  on Knowledge Discovery and Data Mining}, KDD '16, page 855–864, New York,
  NY, USA, 2016. Association for Computing Machinery.
\newblock ISBN 9781450342322.
\newblock \doi{10.1145/2939672.2939754}.
\newblock URL \url{https://doi.org/10.1145/2939672.2939754}.

\bibitem[Kornblith et~al.(2019)Kornblith, Shlens, and Le]{Kornblith_cvpr_2019}
Simon Kornblith, Jonathon Shlens, and Quoc~V Le.
\newblock Do better imagenet models transfer better?
\newblock In \emph{Proceedings of the IEEE Conference on Computer Vision and
  Pattern Recognition}, pages 2661--2671, 2019.

\bibitem[Synerise(2021)]{cleoracode}
Synerise.
\newblock {Cleora code}, 2021.
\newblock URL \url{https://github.com/Synerise/cleora/releases}.

\end{thebibliography}

\end{document}